\documentclass[12pt]{article}
\usepackage{amsfonts}

\begin{document}

\title{\bf The Friedrichs-Model with fermion-boson couplings II.}

\author{
{O. Civitarese$^1$
\thanks{e-mail: civitare@fisica.unlp.edu.ar},
M. Gadella $^2$ \thanks{e-mail: gadella@fta.uva.es},  and G. P.
Pronko$^3$
\thanks{e-mail: pronko@ihep.ru}},\\  \\
{\small\it$^{1}$ Departamento de F\'{\i}sica, Universidad Nacional
de La Plata, }
\\ {\small\it c.c. 67 1900, La Plata, Argentina. } \\
\\ {\small\it$^{2}$Departamento de F\'{\i}sica Te\'orica, Facultad de Ciencias,
E-47011, Valladolid, Spain} \\
\\ {\small\it$^{3}$Institute for High Energy Physics, Protvino,
142284,Moscow Region. Russia}\\ {\small\it and  Institute of Nuclear
Physics, N.C.S. ``Demokritos". } $\;\;\;\;\;\;\;\;\;\;$\\ }

\maketitle

\noindent {\small {\bf Abstract}: In this work we present a formal
solution of the extended version of the Friedrichs Model. The
Hamiltonian consists of discrete and continuum bosonic states,
which are coupled to fermions. The simultaneous treatment of the
couplings of the fermions with the discrete and continuous sectors
of the bosonic degrees of freedom leads to a system of coupled
equations, whose solutions are found by applying standard methods
of representation of bound and resonant states.}
\\


\section{Introduction}

The Friedrichs model  \cite{1} describes the coupling of discrete
and continuum sectors of a boson Hamiltonian. The model has the
advantage of being exactly solvable and it has been applied to a
variety of systems of physical interest \cite{2}. The mathematical
aspects of the solutions which are based in the use of rigged
Hilbert spaces have been introduced in \cite{3}. A distinctive
feature
 of the model is the appearance, among the solutions, of Gamow states
\cite{4,5,6}. It is the only completely solvable model which
describes resonance phenomena \cite{3} in a broad context. Generally
speaking, the solutions show the importance of the
discrete-continuum couplings, a fact that has been overlooked for
quite a while and that certainly has crucial physical consequences.
In fact, the need for the inclusion of bound as well as resonant
states in the description of composite systems has been demonstrated
(see \cite{7} and references quoted therein) in connection with the
analysis of data obtained in experiments measuring the decay of
clusters. The properties of the standard version of the Friedrichs
model, in connection with Gamow resonances, have been revisited in
\cite{5}. More sophisticated versions of the Friedrichs model have
been presented in \cite{6,8,10}. Along the standard
discrete-continuous couplings for bosons, one has to consider also
the boson-boson couplings \cite{2} and fermi-boson couplings
\cite{6}.
 In a recent paper \cite{6}, we
have proposed an extended version of the model to accommodate for
fermion-boson couplings. The structure of the extended Hamiltonian
is inspired in standard field theory models of the coupling between
fermion and boson fields \cite{11,12}. In a previous publication
\cite{6}, hereafter referred to as I, the feasibility of the
extended model was discussed and its solution were explored within a
schematic model situation. By considering a very reduced space
consisting of one discrete fermion state in addition to the bosonic
degrees of freedom, it was found a resonant behavior also in the
fermion sector, as expected. In this paper, we focus our attention
in the analysis of the solution within a more general framework. We
base our method on the use of the resolvent for the characteristic
coupled equations which describes the amplitude of the fermionic
solution. The coupled equations are cast in a form which very much
resembles the T-matrix formalism \cite{13}, which is very convenient
for finding out the resonance behavior.

The present paper is organized as follows: In section 2, we review
the essentials of the extended Friedrichs model, which in full
detail has been introduced in \cite{6}, and focus our attention on
the elements needed for the present discussion. In section 3, we
analyze the structure of the general solution which has been found
by applying the T-matrix formalism. Finally, we draw our conclusions
on section 4.

\section{Formalism}\label{s1}

We shall begin with a review of the basic ideas of the Friedrichs
Model \cite{1,5}. Details about the formalism concerning the
standard Friedrichs model can be found in \cite{AGPP}-\cite{AGP} and
references quoted therein.

The simplest form of the Friedrichs model \cite{1} includes a free
Hamiltonian $H_0$ with a real positive continuous spectrum
coinciding with the positive semiaxis, $\omega
> 0$, and a positive discrete eigenvalue, $\omega_0 >0$. An
interaction is acting between the continuous and discrete parts of
$H_0$ by means of a potential $V$. As a result of the action of $V$,
the bound state of $H_0$ is dissolved in the continuous and a
resonance is produced. The spectrum of the total Hamiltonian $H=H_0+
\lambda V$, where $\lambda$ is a real coupling constant, is purely
continuous and coincides with the real positive semiaxis as well
\cite{1,3}.

As shown in I, poles of the solutions appear into complex conjugate
pairs, each of them represents a resonance. Then, Gamow vectors are
solutions of the eigenvalue equation \cite{2,4},

\begin{equation}
(H-x)\,\Psi(x)=0\,. \label{1}
\end{equation}
with the total Hamiltonian $H$, whose eigenvalues coincide with
resonance poles \cite{3,AGP}.

\subsection{The extended Friedrichs Model.}\label{ss21}

The present version of the Friedrichs model, as an extension of the
standard Friedrichs model, includes \cite{6}:

i) The unperturbed fermion and boson Hamiltonian $H_I$

\begin{eqnarray}
H_{I}=\omega_0|1\rangle \langle 1| + \int_0^\infty d\omega \omega
|\omega \rangle \langle \omega | +\sum_k c_k |k\rangle \langle
k|\,,\label{2}
\end{eqnarray}
where the index $k$ runs out the set of Fermion kets $|k\rangle$.

ii) The interaction between fermions, $|k\rangle$, and the discrete
boson, $|1\rangle$, Hamiltonian $H_{II}$:

\begin{equation}\label{3}
H_{II}=\sum_{k,l}\left[h_{k,l}|k,1\rangle \langle l|+
h^*_{k,l}|l\rangle \langle k,1|\right]\,.
\end{equation}

iii) The interaction between fermions and the boson field,
$\{|\omega\rangle\}$, $H_{III}$:

\begin{equation}\label{4}
H_{III}=\sum_{k,l}\int_0^\infty d\omega \left[f_{k,l}(\omega)
|k,\omega \rangle\langle l|+ f^*_{k,l}(\omega) |l\rangle\langle
k,\omega|\right]\,.
\end{equation}
The standard Friedrichs model includes the boson-boson coupling $V$
\cite{3}. This coupling $V$ can be generalized to include
fermion-boson interactions in the following manner:

\begin{equation}\label{5}
H_{IV}=\sum_{k,k'}\int_0^\infty d\omega \left[g_{kk'}(\omega) |k,1
\rangle\langle k',\omega|+ g^*_{kk'}(\omega)
|k',\omega\rangle\langle k,1|\right]\,.
\end{equation}
Alternatively, $H_{IV}$ can be taken as a second order interaction
mediated by the fermion state $|k\rangle$, provided that:

\begin{equation}\label{6}
\sum_l f^*_{kl}(\omega) h_{lk'}=g_{kk'}(\omega)\,\delta_{kk'}=
g_k(\omega)\,\delta_{kk'}\,, \label{24}
\end{equation}
is fulfilled, as it is customarily done in the field theoretical
treatment of fermion-boson couplings \cite{17}. In addition, it is
worthy to note that $H_{IV}$ reduces to the standard Friedrichs
model interaction, $V$, if $g_{k}(\omega)=g(\omega)$ for all $k$.

To obtain the solution of the eigenvalue problem

\begin{equation}
(H-E)\Psi(E)=0\,,\label{7}
\end{equation}
where $H= H_I+H_{II}+H_{III}+H_{IV}$, we write $\Psi(E)$ in the form

\begin{equation}\label{8}
\Psi(E)=\sum_k\varphi_k(E)|k\rangle+\sum_k \phi_{k,1}(E)|k,1\rangle
+\sum_k\int_0^\infty d\omega \,\psi_k(E,\omega)|k,\omega\rangle\,.
\end{equation}
Consequently, (\ref{7}) gives

\begin{eqnarray}
&&(H-E)\Psi(E)\equiv \sum_k\varphi_k(E)(c_k-E)|l\rangle
+\sum_{k,l} h_{kl}^*\phi_{k,1}(E)|k\rangle \nonumber \\
&+&\sum_k (c_k+\omega_0-E)\phi_{k,1}(E)|k,1\rangle
+\sum_k\int_0^\infty d\omega \psi_k(E,\omega)g_k(\omega)|k,1\rangle
\nonumber
\\ &+&\sum_k \phi_{k,1}(E)\int_0^\infty d\omega
g_k^*(\omega)|k,\omega\rangle +\sum_k\int_0^\infty d\omega
\psi_k(E,\omega)(c_k+\omega-E)|k,\omega\rangle\nonumber
\\ &+&\sum_{k,l}\int_0^\infty d\omega
\psi_k(E,\omega)f_{kl}^*(\omega)|l\rangle
+\sum_{i,k}\varphi_k(E)h_{ik}|i,1\rangle\nonumber \\
&+&\sum_{i,k}\int_0^\infty d\omega
\varphi_k(E)f_{ik}(\omega)|i,\omega\rangle=0\,. \label{9}
\end{eqnarray}
Due to the linear independence of the vectors $|k\rangle$,
$|k,1\rangle$, and $|k,\omega\rangle$, the above equation can be
written as a system of the following three coupled equations:

\begin{equation}
\varphi_k(E)(c_k-E)+\sum_lh_{lk}^*\phi_{l1}(E)+\sum_l\int_0^\infty
d\omega \psi_l(E,\omega)f_{lk}^*(\omega)=0 \label{10}
\end{equation}

\begin{equation}
\sum_l\varphi_l(E)h_{kl}+(c_k+\omega_0-E)\phi_{k1}(E)+\int_0^\infty
d \omega \psi_k(E,\omega)g_k(\omega)=0 \label{11}
\end{equation}

\begin{equation}
\psi_k(E,\omega)(c_k+\omega-E)+\sum_l\varphi_l(E)f_{kl}
(\omega)+\phi_{k1}(E)g_k^*(\omega)=0 \label{12}
\end{equation}
for $|k\rangle$, $|k,1\rangle$, and $|k,\omega\rangle$,
respectively. using (\ref{10}) and (\ref{11}) in (\ref{12}),  we
obtain the following equation:

\begin{equation}
\psi_k(E,\omega)=c\delta(c_k-\omega-E)- \sum_l\frac{ \varphi_l(E)
f_{kl}(\omega)}{c_k-\omega-E}-\frac{\phi_{k1}(E)g_k^*(\omega)}{c_k-\omega-E}\,.
\label{13}
\end{equation}
Consequently, if $\psi_k(E,\omega)$ is reinserted into (\ref{10})
and (\ref{11}), it results a system of infinite coupled equations.
The formal properties of its solutions becomes obvious at glance:
i.) the resonant behavior of the fermionic solutions results from
the coupling to the continuum-continuum boson sector; ii.) the
resonant boson sector is mostly unaffected by the fermion sector,
since the latter solely depends on $V$.

The solutions of the system depend on the choices of the form
factors $h_{kl}$, $f_{kl}(\omega)$ and $g_k(\omega)$. After a
rearrangement, (\ref{10}) and (\ref{11}) yields:

\begin{eqnarray}
\left[(c_k-E)\delta_{km}- \sum_m A_{km}(E)\right]
\varphi_m(E)\nonumber\\[2ex]+\sum_m\left(
h_{mk}^*\right. -\left. B_{km}(E)\right)\phi_{m1}(E)=-c\sum_m
f_{mk}^*(E-c_m)\label{14}
\end{eqnarray}
and

\begin{eqnarray}
&&\sum_l [h_{kl}-\widetilde B_{kl}(E)]\varphi_l(E)\nonumber\\[2ex]&&+
(c_k+\omega_0-E-C_k(E))\phi_{k1}(E)=-cg_k(E-c_k)\,.\label{15}
\end{eqnarray}

Note that the functions, $A_{km}(E)$, $B_{km}(E)$, $\widetilde
B_{kl}(E)$ and $C_k(E)$ are explicit functions of the form factors.
These functions have been given in I, (see Equations 32-35 of
\cite{6}). The factor $c$ in (\ref{15}) is an irrelevant constant.

\section{Choice of form factors.}

In I, we have shown that the system of coupled equations which
determine the amplitudes $\varphi_k(E)$, $\phi_{l1}(E)$ and
$\psi_k(E,\omega)$ can be solved under certain simplifications like
the assumption that the form factors do not depend on the fermionic
indices and/or that the fermionic subspace be severely reduced.
Here, we propose a more general setting for finding the solution,
which consists of the study of the separate couplings of the
fermions with the discrete and continuous bosons. For the form
factors, we here propose the following choice:

\begin{equation}
 f_{lk}(\omega)=\alpha_l\alpha_k^* \rho(\omega) \,,
\hskip1cm h_{lk}=\beta_l\beta_k^*\,,\hskip1cm g_k(\omega)=0\,,
\label{16}
\end{equation}
This choice does not contradict (\ref{6}) because we are considering
the couplings (\ref{2}), (\ref{3}) and (\ref{4}) as independent
couplings, of which (\ref{6}) is a special case only. Here,
$\rho(\omega)$ is  an unspecified function, whose explicit structure
is not needed for the present discussion. In actual calculations it
is defined by the model Hamiltonian.  Thus, with this choice of the
form factors, Equation (\ref{15}) reads:

\begin{equation}\label{17}
\phi_{k1}(E)= -\sum_s \frac{\varphi_s(E)\,h_{ks}}{c_k-\omega_0-E}
\,. \label{17}
\end{equation}

Then, Equations (32-35) in I  become

\begin{eqnarray}
A_{km}(E)=\int_0^\infty d\omega\sum_l
\frac{\alpha_l^*\alpha_k\alpha_l\alpha_m^*\,|\rho|^2}{c_l+\omega-E}\nonumber\\[2ex]
=\alpha_k\alpha_m^* \int_0^\infty d\omega\sum_l
\frac{\alpha_l^*\alpha_l\,|\rho|^2}{c_l+\omega-E}=
\alpha_k\alpha_m^*\,A(E)\,.\label{18}\\[2ex]
B_{km}=0\,,\hskip1cm \widetilde B_{km}=0 \,,\hskip1cm
C_{km}=0\,.\label{19}
\end{eqnarray}
Note that (\ref{18}) defines $A(E)$. This yields the following form
for (\ref{14}):

\begin{eqnarray}
\sum_m\left[(c_k-E)\,\delta_{km}-\alpha_k\alpha_m^*\,A(E)\right]\,\varphi_m(E)-\sum_lB(E)\,\beta_k\beta_l^*
\varphi_l\nonumber\\[2ex] =-c\sum_m f_{mk}^*(E-c_m)\,,\label{20}
\end{eqnarray}
where  $B(E)$ is

\begin{equation}\label{21}
B(E)=\sum_m\frac{\beta_m^*\beta_m}{c_m+\omega_0-E}\,.
\end{equation}
Equation (\ref{20}) can be written in a more compact form as

\begin{equation}
\sum_s\varphi_s(E)[(c_k-E)\delta_{ks}-\alpha_k^*\alpha_s\, A(E)-
\beta_k^*\beta_s\, B(E)] = -c\sum_l f_{lk}^*(c_l-E)\,, \label{22}
\end{equation}
where the functions on the variable $E$, $A(E)$ and $B(E)$, do not
depend either on $k$ and $s$. Equation (\ref{22}) can be also
written in matrix form:

\begin{equation}
T\left(\begin{array}{c} \cdots\\ \varphi_s(E)\\ \cdots
\end{array}\right)= \left(\begin{array}{c} \cdots\\ -c\sum_l f_{lk}^*(c_l-E)\\ \cdots
\end{array}\right)\,, \label{23}
\end{equation}
where the entries in the matrix $T$ are given by

\begin{eqnarray}
T_{ks}= (c_k-E)\delta_{ks}-\alpha_k^*\alpha_s\, A(E)-
\beta_k^*\beta_s\, B(E) \nonumber\\ [2ex] =
(c_k-E)\left\{\delta_{ks}-\frac{\alpha_k^*\alpha_s\, A(E)}{(c_k-E)}
- \frac{\beta_k^*\beta_s\, B(E)}{(c_k-E)} \right\}\,. \label{24}
\end{eqnarray}
To some extent, we have borrowed the notion frequently encountered
in Theory of Reactions (see Newton \cite{13}) in dealing with the
formal solution of systems of coupled equations, where $T$ stands
for the transfer matrix. A solution to our problem with the choice
(\ref{16}) can be obtained provided that T be invertible. To prove
this claim, we use the matrix identity $(1-A)^{-1}=\sum_n A^n$. In
consequence, the entries of $T^{-1}$ are the following:

\begin{eqnarray}
(T^{-1})_{ks}= \frac{1}{c_k-E}\left\{\delta_{ks}-
\frac{\alpha_k^*\alpha_s\, A(E)}{c_k-E} -\frac{\beta_k^*\beta_s\,
B(E)}{c_k-E} \right\}^{-1} \nonumber\\ [2ex] =
\frac{1}{c_k-E}\left\{\delta_{ks} +\left( \frac{\alpha_k^*\alpha_s\,
A(E)}{c_k-E} + \frac{\beta_k^*\beta_s\, B(E)}{c_k-E}
\right)\right. \nonumber\\[2ex]+ \left.\left( \frac{\alpha_k^*\alpha_{k'}\,
A(E)}{c_k-E} + \frac{\beta_k\beta_{k'}\, B(E)}{c_k-E} \right) \left(
\frac{\alpha_{k'}^*\alpha_s\, A(E)}{c_{k'}-E} +
\frac{\beta_{k'}\beta_s\, B(E)}{c_{k'}-E}
\right) +\cdots \right\}\nonumber\\[2ex]
= \frac{\delta_{ks}}{c_k-E} + \alpha^*_k\alpha_s a_{ks}+
\beta^*_k\beta_s b_{ks}+ \alpha^*_k\beta_s c_{ks}+ \beta_k^*\alpha_s
d_{ks}\,. \label{25}
\end{eqnarray}
Then, it remains to determine the coefficients $a_{ks}$, $b_{ks}$,
$c_{ks}$, $d_{ks}$, with the use of the obvious relation $\sum_{k'}
T_{kk'}^{-1}T_{k's}=\delta_{ks}$, which gives a  expression of the
form:

\begin{eqnarray}
\alpha^*_k\alpha_s C_1(a_{ks},b_{ks},\dots)+ \beta^*_k\beta_s
C_2(a_{ks},b_{ks},\dots)
 &+&\alpha^*_k\beta_s C_3(a_{ks},b_{ks},\dots)  \nonumber\\[2ex] &+&
\beta_k^*\alpha_s C_4(a_{ks},b_{ks},\dots)=0\,. \nonumber\\
\label{26}
\end{eqnarray}
If we assume linear independence of the variables $\alpha_k$ and
$\beta_s$, the coefficients of (\ref{26}) should vanish for all
values of the indices $r$ and $s$:

$$
C_p(a_{ks},b_{ks},\dots)=0\,,\hskip2cm p=1,2,3,4\,.
$$
By implementing in (\ref{25}) condition (\ref{26}) one obtains

\begin{eqnarray}
(c_s-E) a_{ks}-\frac{A(E)}{c_k-E} -A(E)\sum_{k'}[
\alpha_{k'}\alpha_{k'}^* a_{k's} +\beta_{k'}\alpha_{k'}^* c_{kk'}]
=0 \,. \label{27}
\\ [2ex] (c_s-E) b_{ks}-\frac{B(E)}{c_k-E} -B
(E)\sum_{k'}\left[\beta_{k'}\beta_{k}^* b_{kk'} +
\alpha_{k'}\beta_{k'}^* d_{kk'}\right]=0\,. \label{28} \\[2ex]
c_{ks}(c_s-E)-B(E)\sum_{k'}[\alpha_{k'}\beta_{k'}^*
a_{kk'}+\beta_{k'}\beta_{k'}^* c_{kk'}]=0\,.
\label{29} \\[2ex]
d_{ks}(c_s-E) -A(E)\sum_{k'}[\beta_{k'}\alpha_{k'}^* b_{kk'}+
\alpha_{k'}\alpha_{k'}^* d_{kk'}]=0\,. \label{30}
\end{eqnarray}
Let us start by solving Equation (\ref{29}). It can be written as

\begin{equation}
c_{ks}= \frac{B(E)R_k}{c_s-E} \,, \label{31}
\end{equation}
where

\begin{equation}
R_k= \sum_{k'}[\alpha_{k'}\beta_{k'}^*
a_{kk'}+\beta_{k'}\beta_{k'}^* c_{kk'}]\,. \label{32}
\end{equation}
In order to simplify subsequent equations, we shall introduce the
following notation:

\begin{equation}\label{33}
\tau(\alpha,\beta,E):= \sum_k\frac{\alpha_k\beta_k}{c_k-E}\,.
\end{equation}

Using this definition in (\ref{31}-\ref{32}) gives

\begin{equation}\label{34}
\frac{B(E)\,R_k}{c_s-E}=\frac{B(E)}{c_s-E}\left(\sum_{k'}\alpha_{k'}\beta_{k'}^*
a_{kk'}+\tau(\beta,\beta^*,E)\,B(E)\,R_k\right)\,,
\end{equation}
and from it the quantity $R_k$ can be expressed as

\begin{equation}
R_k=\frac{\sum_{k'} a_{k'}\beta_{k'}^*
a_{kk'}}{1-B(E)\,\tau(\beta,\beta^*,E)}\,.\label{35}
\end{equation}
In a similar manner

\begin{equation}
c_{ks}=\frac{B(E)}{c_s-E}\,\sum_{k'}\frac{\alpha_{k'}\beta_{k'}^*a_{kk'}}{1-B(E)\,\tau(\beta,\beta^*,E)}\,.\label{36}
\end{equation}
and

\begin{equation}
a_{ks}=\frac{A(E)}{c_s-E}\,\left[\frac{1}{c_k-E}+
\sum_{k'}(\alpha_{k'}\alpha_{k'}^*a_{kk'}+\beta_{k'}\alpha_{k'}c_{kk'})\right]\,.\label{37}
\end{equation}
Note that the inverse matrix $T^{-1}$ does exist if and only if the
$R_k$ are nonsingular for all $E$, i.e., the denominator in
(\ref{35}) does not vanish:

\begin{equation}\label{38}
    1-B(E)\,\tau(\beta,\beta^*,E)\ne 0\,.
\end{equation}
If one proceeds with the expression of $a_{ks}$ in (\ref{27}) as
done with $c_{ks}$, one finds

\begin{equation}
a_{ks}=\frac{A(E)\,\Pi_k}{c_s-E}\,,\label{39}
\end{equation}
with

\begin{equation}
\Pi_k:=
\frac{1}{c_k-E}+\sum_{k'}\alpha_{k'}\alpha_{k'}^*a_{kk'}+\frac{B(E)\,\tau(\beta,\alpha^*,E)}{1-B(E)\,
\tau(\beta,\beta^*,E)}\,\sum_{k'}\alpha_{k'}\beta_{k'}^*a_{kk'}\,.\label{40}
\end{equation}
From (\ref{39}) and (\ref{40}), one gets

\begin{eqnarray}
\frac{A(E)\,\Pi_k}{c_s-E}=\frac{A(E)}{c_s-E}\left[\frac{1}{c_k-E}+A(E)\Pi_k\,
\tau(\alpha,\alpha^*,E)\right.\nonumber\\[2ex]
\left.+
B(E)\,\frac{\tau(\beta,\alpha^*,E)\,\tau(\alpha,\beta^*,E)}{1-\tau(\beta,\beta^*,E)\,B(E)}
+A(E)\Pi_k\right]\,, \label{41}
\end{eqnarray}
from which we obtain a consistent equation for $\Pi_k$:

\begin{eqnarray}
\Pi_k=\frac{1}{c_k-E}\left[1-A(E)\,\tau(\alpha,\alpha^*,E)-A(E)B(E)\,\frac{\tau(\beta,\alpha^*,E)\,
\tau(\alpha,\beta^*,E)}{1-\tau(\beta,\beta^*,E)\,B(E)}\right]\,,\label{42}
\end{eqnarray}
leading to

\begin{equation}
a_{ks}=\frac{A(E)\,(1-B(E)\,\tau(\beta,\beta^*,E))}{(c_k-E)(c_s-E)}\,\Delta^{-1}\,,\label{43}
\end{equation}
with

\begin{eqnarray}
\Delta:=
1-A(E)\,\tau(\alpha,\alpha^*,E)-B(E)\,\tau(\beta,\beta^*,E)\nonumber\\[2ex]+
A(E)B(E)\,[\tau(\alpha,\alpha^*,E)\,\tau(\beta,\beta^*,E)-\tau(\alpha,\beta^*,E)\,\tau(\beta,\alpha^*,E)]\,.\label{44}
\end{eqnarray}
Accordingly, nontrivial solutions of $a_{ks}$ ($a_{ks}\ne 0$) exist
if and only if $\Delta\ne 0$ for all values of the energy $E$.

Finally, the expressions for the remaining coefficients are

\begin{equation}
b_{ks}=\frac{B(E)\,(1-A(E)\,\tau(\alpha,\alpha^*,E))}{(c_s-E)(c_k-E)}\,\Delta^{-1}\,,
\label{45}
\end{equation}

\begin{equation}
d_{ks}=
\frac{A(E)B(E)\,\tau(\beta,\alpha^*,E)}{(c_s-E)(c_k-E)}\,\Delta^{-1}\,.
\label{46}
\end{equation}
With these elements, we are now in the position of writing the
explicit form for the inverse matrix $T^{-1}$

\begin{eqnarray}
T_{ks}^{-1}=\frac{\delta_{ks}}{c_k-E}+\frac{\Delta^{-1}}{(c_k-E)(c_s-E)}
\,\left\{\alpha_k^*\alpha_s\,A(E)\,[1-B(E)\,\tau(\beta,\beta^*,E)]\right.\nonumber
\\[2ex] +\beta_k^*\beta_s
\,B(E)\,[1-A(E)\,\tau(\alpha,\alpha^*,E)]+\alpha_k^*\beta_s^*\,A(E)\,B(E)\,\tau(\alpha,\beta^*,E)
\nonumber\\[2ex]
+\beta_s^*\alpha_s\,A(E)\,B(E)\,\tau(\beta,\alpha^*,E)\}\,.
\label{47}
\end{eqnarray}
In consequence, the formal solution of Equation (\ref{23}) is given
by

\begin{equation}
\varphi_k(E)=-c\sum_k T^{-1}_{kn}\left(\sum_l f_{ln}^*(c_l-E)\right)
\label{48}
\end{equation}
and

\begin{equation}
\phi_{k1}(E)=c\frac{\beta_k}{c_k+\omega_0-E}\sum_{n,m}\beta^*_m\,T^{-1}_{mn}\left(\sum_l
f_{lk}^*(c_l-E)\right)\,.\label{49}
\end{equation}
Thus, inserting (\ref{48}) and (\ref{49}) in (\ref{13}) and then in
(\ref{8}) the solution to our problem is completely determined. Note
that the structure of the solutions is preserved by form factors
which may not been separable as those of the choice (\ref{16}),
since they are absorbed in the definition of the functions $\tau$.

\section{Concluding remarks.}

In this paper, we have described the solutions of the extended
Friedrichs model. The proposed model includes fermion-boson
couplings in addition to the boson-boson couplings. From the
physical point of view, this extension may become a suitable tool
for the treatment of decaying states which participate in the
observable total decay width of composite nuclear systems. From the
mathematical point of view, we have generalized the schematic
solution proposed in a previous article \cite{6} and formulated the
solution in a more compact and formal way. The main features of the
proposed method can be summarized as follows:

i.) The resonant behavior of the fermionic sector depends on the
resonant structure of the boson field, which participate in the
fermion solution via the fermion-boson coupling.

ii.) The set of coupled equations is greatly simplified by the use
of a T-matrix like formalism, which allows us to identify the
resonant behavior of the solutions.

Further applications of the set of Equations (\ref{47}-\ref{49}) to
situations of physical interest are in progress.

\section*{Acknowledgements.}

Financial support is acknowledged to the Ministry of Education of
Spain, grants PR2004-0080, MTM2005-09183, FIS2005-03989,
SAB2004-0169, the Junta de Castilla y Le\'on Project VA013C05, the
Russian Science Foundation Grant 04-01-00352 and the CONICET
(Argentina).

\end{document}